# Reconstructing Positions and Peculiar Velocities of Galaxy Clusters within 20000 km/sec. I: The Cluster 3D Dipole

Enzo Branchini[1] & Manolis Plionis[1,2]

[1] *SISSA - International School for Advanced Studies, Via Beirut 2-4, I-34013 Trieste, Italy*
[2] *National Observatory of Athens - Astronomical Institute, Lofos Nimfon, Thesio, 11810, Athens, Greece*

12 January 1995

**ABSTRACT**
Starting from the observed distribution of galaxy clusters in redshift space we use a two–step procedure to recover their distances and peculiar velocities. After statistically correcting for the unobserved cluster distribution in the zone of avoidance ($|b| \leq 20°$) and also for a smooth absorption at higher $|b|$'s, we use a dynamical iterative algorithm to recover the real–space cluster positions by minimizing the redshift space distortions. The whole procedure assumes that clusters trace the mass, that peculiar velocities are caused by gravity and that linear perturbation theory applies. The amplitude of the cluster dipole measured in the 3D space turns out to be $\sim 23\%$ less than that measured in redshift space. In both cases the dipole direction is aligned with the Cosmic Microwave Background dipole within $\sim 10°$, taking into account the Virgocentric infall component of the Local Group [LG hereafter] motion. Observational errors, limitations in the reconstruction procedure and the intrinsic cosmological variance, which is the dominant source of uncertainty, render a stringent determination of the $\beta (\equiv \Omega_0^{0.6}/b)$ parameter whose central value turns out to be $\beta \approx 0.2$ while its total uncertainty is $\pm 0.1$. This implies that for a cluster-mass bias parameter of $\sim 5$, a flat Universe is not excluded, contrary to previous cluster-dipole $z$-space analysis. A more stringent determination of $\beta$ will be obtained from the analysis of the peculiar velocity field in a forthcoming paper.

## 1 INTRODUCTION

The study of cosmic peculiar velocities is a relatively recent and very exciting field that could give great insight to the origin of the large-scale structures in the Universe. With the advent of new observational techniques and telescope power it has become possible to relate distance dependent with distance independent quantities and thus determine the relative distance for a large number of galaxies; the most frequently of these being used are the *Tully-Fisher* relation for spirals and the $D_n - \sigma$ relation for elliptical galaxies (cf. Burstein 1990 and Dekel 1994 for comprehensive reviews).

The best evidence that cosmic structures can have significant velocities, above their cosmological expansion velocities, came from the interpretation of the CMB temperature dipole anisotropy as a Doppler effect, originating from the motion, with $v_{LG} = 622 \pm 20$ km/sec, of the Local Group of galaxies in the isotropic CMB radiation sea, towards $l = 277°$ and $b = 30°$ (Lubin & Villela 1986; Kogut et al. 1993)

If linear gravity is responsible for the observed peculiar velocities then according to the linear instability theory (cf. Peebles 1980) the peculiar gravitational acceleration should be aligned and proportional to the peculiar velocity; the constant of proportionality being a measure of the present-time growth rate of mass fluctuations and therefore a measure of the cosmological density parameter, $\Omega_0$. The gravitational acceleration of any galaxy (or other extragalactic object) can be estimated by calculating the dipole moment of the distribution of mass surrounding it. Since, however, the latter is unknown one has to resort in estimating the dipole moment of the distribution of luminous extragalactic objects using some simplifying assumption about the relation between fluctuations in the matter and light distributions; which is non other than the *linear-biasing* assumption (cf. Kaiser 1984). In this picture the galaxy fluctuations, $\delta_g$, are related to those of the matter distribution $\delta$ by a constant factor, the biasing parameter, $\delta_g = b\delta$.

Further difficulties arise from the fact that one can observe galaxies or other extragalactic objects limited in magnitude or flux, which then implies that distant contributions to the dipole could be missed if the characteristic depth of the galaxy catalogue is less than the convergence depth of the dipole. Moreover, most catalogues of extragalactic objects have also limited sky coverage which is usually due to light absorption near the Galactic plane.

Up to now the dipole of various populations of extragalactic objects has been determined: optical galaxies (Lahav 1987; Plionis 1988; Lahav, Rowan-Robinson & Lynden-Bell 1988; Lynden-Bell, Lahav & Burstein 1989; Hudson 1993b), IRAS galaxies (Meiksin & Davis 1986; Yahil, Walker & Rowan-Robinson 1986; Villumsen & Strauss 1987; Strauss & Davis 1988; Strauss et al. 1992; Rowan-Robinson et al. 1990; Plionis, Coles & Catelan 1993), X-ray active galactic nuclei (Miyaji & Boldt 1990); X-ray clusters (Lahav et al. 1989) and Abell clusters (Scaramella, Vettolani & Zamorani 1991 [hereafter SVZ91]; Plionis & Valdarnini 1991 [hereafter PV91]). In all cases the dipole moment was found to be quite well aligned with the CMB dipole suggesting that gravity is indeed responsible for the Local Group motion and that light traces mass.

An interesting historical fact, originating from these studies, is that the estimated dipole convergence depth has been a function of cosmic volume sampled; *the deeper the catalogue the larger the dipole convergence depth*. This is true out to the largest depths, traced by the Abell/ACO cluster catalogue, for which $R_{conv} \approx 17000$ km/sec (PV91 and SVZ91). This implies that the apparent dipole convergence of the shallower galaxy catalogues (being optical or IRAS) is probably artificial, determined by their limiting depth. In fact a careful investigation of the QDOT-IRAS sample showed some evidence for a contribution to the dipole from scales comparable to the cluster $R_{conv}$ (Plionis, Coles & Catelan 1993).

A further interesting outcome of the Abell/ACO cluster dipole analysis is the fact that there seems to be a *coherent* dipole anisotropy out to $\sim 17000$ km/sec; i.e., the differential gravitational acceleration induced on the LG by the distribution of clusters in large equal-volume shells is roughly aligned with the CMB dipole in each shell (PV91). Furthermore, there is strong evidence for the existence of coherent large-scale galaxy flows in the local Universe, extending from the Perseus-Pisces region on the one side to the Hydra-Centaurus/Great Attractor region on the other. These flows are well established both for elliptical (cf. Lynden-Bell et al. 1988) and spiral galaxies (cf. Rubin et al. 1976; Dressler & Faber 1990; Mould et al. 1991; Willick 1990; Han & Mould 1990; Courteau et al. 1993; Hudson 1994; Mathewson & Ford 1994) and point in the general direction of the CMB dipole. These results put together present a consistent picture in which the LG participates in a large-scale bulk motion induced by gravity, encompassing a large volume of radius $\sim 6000 - 15000$ km/sec (see the review of Dekel 1994). Although the extent of this volume may still be under debate the existence of the bulk flow seems to be certain.

This picture has been recently challenged by Lauer & Postman (1994) [hereafter LP94] who have extended the cosmic flow studies to very large scales, those traced by galaxy clusters, and who find that the LG motion with respect to the frame defined by Abell/ACO clusters, within 15000 km/sec, moves in a direction very different than that of the CMB dipole which then implies that, if the CMB dipole is a Doppler effect, the whole cluster frame is moving with respect to the CMB rest-frame with $\sim 700$ km/sec. Such a large velocity on such large scales has put into despair structure formation model-builders, since no model can easily accommodate such velocities on large scales (cf. Strauss et al. 1994; Feldman & Watkins 1994; Jaffe & Kaiser 1994).

Furthermore and more importantly it puts into doubt the gravitational instability picture because on the scales traced by clusters of galaxies linear theory should apply, which predicts that acceleration and velocity are aligned and as discussed previously, the gravitational acceleration of the LG of galaxies, determined from the cluster distribution in $z$-space and within $\sim 25000$ km/sec, is found to be very well aligned ($\delta\theta_{cmb} \leq 15°$) with the CMB dipole (PV91; SVZ91).

So we are left with the following puzzling picture: the acceleration determined from the distribution of galaxies (within $\sim 10000$ km/sec) and of clusters (within $\sim 25000$ km/sec) is well aligned with the CMB dipole (which is in itself a strong indication that the CMB dipole is Doppler generated) while the LG velocity (as determined by LP94), with respect to the cluster frame within 15000 km/sec, points to a direction almost perpendicular to the CMB dipole direction. One would then be forced to explain the alignment of the LG gravitational acceleration, determined from the distribution of clusters and galaxies, with the CMB dipole direction as a *'cosmic'* coincidence. However, the random joint probability of having the cluster dipole, the optical and IRAS galaxy dipoles aligned with the CMB dipole within a few degrees ($\sim 10°$, $\sim 10°$ and $\sim 25°$ respectively), assuming that they are independent, is $\leq 3 \times 10^{-6}$. Furthermore, as we discussed previously, PV91 found evidence for a coherent dipole anisotropy, in which the differential cluster dipole in each equal-volume shell roughly points towards the CMB dipole. In fact, the dipole in the first shell ($R \lesssim 100\ h^{-1}$ Mpc) has $\delta\theta \lesssim 20°$ while in the most distant shell ($140 \lesssim R \lesssim 160\ h^{-1}$ Mpc), for which the dipole is aligned with the CMB, it has $\delta\theta \lesssim 10°$ and the distance between the centers of the two shells is $\sim 90 \, h^{-1}$ Mpc, a distance at which $\xi_{cc}(r) \ll 1$, where $\xi_{cc}$ is the spatial cluster–cluster correlation function (note that even the edges of the two shells are $\sim 43 \, h^{-1}$ Mpc away). Therefore, these two volumes could be considered independent which would then decrease even further the probability of random alignment with the CMB dipole to $\lesssim 8 \times 10^{-8}$!

We attempt to throw light to this paradox by using linear perturbation theory and a dynamical reconstruction algorithm (similar to that of Strauss & Davis 1988) to determine the 3D cosmic density field, traced by the Abell/ACO galaxy clusters within $\lesssim 20000$ km/sec, with the aim of investigating

(i) whether the large cluster-dipole convergence depth, its asymptotic amplitude and the alignment of the $z$-space cluster dipole with that of the CMB, are artifacts of redshift space distortions,
(ii) the peculiar velocity field, within 20000 km/sec, predicted by linear theory and comparing it directly with that derived by LP94.

In this first paper we will address the first set of questions, outlined above; while the second will be addressed in a forthcoming paper. The plan of this paper is as follows: In section 2 we present an extended discussion of the two subsamples used, of their biases and homogenization procedure. In section 3 we present both the zone of avoidance *filling* procedure, the dynamical real–space reconstruction algorithm and its reliability tests. In section 4 we present the cluster dipole estimate, possible systematic effects and an extended error analysis. In section 5 we derive interesting cosmological parameters from the dipole analysis and finally in section 6 we present our conclusions.

## 2 THE CLUSTER SAMPLE

Our present analysis is based on a cluster sample extracted from the original Abell sample (Abell 1958) and from its southern extension, the ACO cluster catalogue (Abell, Corwin & Olowin 1989). The north declination limit of the ACO sample is $\delta = -17$, so that the two catalogues overlap in the strip $-27 \leq \delta \leq -17$. The overlapping region has been used by Abell, Corwin & Olowin (1989) to calibrate the $J$ magnitudes of the ACO to the Abell magnitude system. However, systematic differences in the two calibrated catalogues have been noticed by ACO (1989), Batusky et al. (1989), Scaramella et al. (1990) and PV91. The problem of obtaining a statistically homogeneous catalogue from the composite Abell/ACO sample, is particularly important in dipole studies since artificial systematic differences (density gradients or global density variations) can enhance or even produce a dipole signal. Although this issue has been addressed by a number of authors (cf.

Scaramella et al. 1990 and PV91) we will also discuss it, following the PV91 line of reasoning.

Cluster distances, $r$, are estimated from redshifts using the standard relation:

$$r = \frac{c}{H_0 q_0^2 (1+z)} \left[ q_0 z + (1 - q_0)\left(1 - \sqrt{2 q_0 z + 1}\right)\right], \quad (1)$$

where $c$ is the speed of the light and $q_0$ the deceleration parameter. The number of clusters within our subsample is not fixed since the distances slightly depend on $q_0$ which we do not fix *a priori*. We also merge cluster pairs with relative distance $\leq 3 \, h^{-1}$ Mpc into unique 'clusters' having as position their center of mass and as mass their combined total mass. We made this approximations to avoid nonlinear effects on small scales for which our reconstruction method fails to account for. The combined subsample we consider is composed by:

(i) Abell clusters in the roughly volume–limited region $r \leq 250 \, h^{-1}$ Mpc, $|b| \geq 13°$ and with $m_{10} < 17$. $m_{10}$ is the magnitude of the tenth brightest galaxy in the cluster in the magnitude system corrected according to PV91. This sample is 100 % redshift complete. Note that the number of objects in this subsample varies between 269 and 276 for $q_0 = 0.2$ and 0.5, respectively. Out of these we have merged 5 cluster pairs.
(ii) ACO clusters with $r \leq 250 \, h^{-1}$ Mpc, $|b| \geq 13°$ and $m_{10} < 17$. Among the 217 objects within the sample $\sim 77\%$ have measured redshift. For the remaining objects the redshift is determined using the $m_{10} - z$ relation used by PV91. Note that we have merged 7 nearby cluster pairs.

In order to obtain a statistically homogeneous all–sky sample of clusters we quantify separately the observational biases, present in the Abell and ACO subsamples, and then we calibrate these subsamples applying a homogenization procedure.

### 2.1 The Selection Biases

At low galactic latitudes, absorption of optical photons makes it unlikely to observe luminous objects. In the galactic strip $|b| \leq 20°$, to which we refer as the Zone of Avoidance (ZoA), there are only 20 clusters in our subsample. The galactic absorption outside the ZoA has been always found to be consistent with a cosecant law:

$$\log P(b) = \alpha(1 - \text{cosec}|b|), \quad (2)$$

where $P(b)$ gives the probability that a cluster in the range $b$ to $b + \Delta b$ would have been included in the catalogue. There is no general agreement on the precise value of $\alpha$. Bahcall & Soneira (1983) and Postman et al. (1989) found 0.3 for Abell clusters, Batuski et al. (1989) found $\alpha \sim 0.2$ for ACO clusters while LP94 found an even lower value ($\alpha \sim 0.15$) for both ACO and Abell clusters contained in a volume–limited 15000 km/sec sphere. To allow for this experimental uncertainty we will consider, in what follows, two different sets of values: $(\alpha,\alpha)= (0.3,0.2)$, that better matches the behaviour of the number density of our cluster as a function of the galactic latitude and $(0.2,0.2)$. The first value in each set refers to the Abell sample while the second to the ACO one.

The finite depth of the cluster sample was modeled with the redshift selection function that was estimated as in Postman et al. (1989). We modeled the probability that a cluster in the range $z$ to $z + \Delta z$ would have been included in the cluster catalogue as:

$$P(z) = \begin{cases} 1 & \text{if } z \leq z_c \\ A \, \exp(-z/z_0) & \text{if } z > z_c \end{cases} \quad (3)$$

where $z_c$ is the redshift up to which the space density of clusters remains constant (volume-limited regime). We obtain $z_c \sim 0.0787$ and $0.0664$ for the Abell and ACO subsamples respectively.

No appreciable declination dependence has been detected in the Abell/ACO subsample (but see Scaramella et al. 1990 for a possible zenithal dependence).

## 2.2 Homogenization of the Abell/ACO sample

In this work we will follow the PV91 homogenization scheme, used to minimize the density variations of the Abell and ACO subsamples. The estimated observational biases allow us to assign the following weight to clusters:

$$w_i(z,b) = \frac{1}{P(z_i)P(b_i)}. \quad (4)$$

The individual number density of clusters of the two subsamples is therefore $n = V^{-1} \sum_i w_i(z,b)$, where the sum extends over all clusters within the sampled volume $V$. Since at low galactic latitudes the patchy galactic absorption is not well described by eq.(2) we compute $n$ above $|b| = 30°$. We divided $V$ into 15 equal volume intervals in which we estimated the Abell and ACO cluster number density separately. Due to the complex geometric boundaries, the volumes were computed by Montecarlo calculations.

Within $\sim 200\ h^{-1}$ Mpc the space density of ACO and Abell is roughly constant with $n \sim 2.5 \pm 0.3 \times 10^{-5}$ and $1.5 \pm 0.2 \times 10^{-5}\ h^3$ Mpc$^{-3}$, respectively. Errors were computed from the density fluctuations in the different bins. Although beyond $\sim 200\ h^{-1}$ Mpc the density errors of the Abell sample remain roughly the same, those of the ACO increase dramatically ($\sim 10^{-5}\ h^3$ Mpc$^{-3}$) implying large density fluctuations due to undersampling.

An approximate homogenization of the two samples was then done by equating the number density of the two catalogues in each equal volume region. Assuming that the two catalogues are simply radially inhomogeneous, we computed the following weighting function:

$$\mathcal{W}_{rel}(r) = \begin{cases} 1 & \text{if } \delta \geq -17 \\ \frac{n(r_i, Abell)}{n(r_i, ACO)} & \text{if } \delta < -17 \end{cases} \quad (5)$$

where $r_i$ is the distance of the $i^{\text{th}}$ bin. No significant difference was found when using the two different sets of absorption coefficients. Note, however, that this is not the only possible homogenization scheme since we do not know *a priori* which is the true value of the cluster density. We could, therefore, homogenize the two samples by using the inverse of eq.(5), $\mathcal{W}_{rel}(r)^{-1}$. As we will show in section 4 our main results remain stable on the choice of the homogenization scheme.

## 3 THE RECONSTRUCTION METHOD

The whole reconstruction procedure can be schematically represented by the following diagram:

$$z_{obs} \longrightarrow z_{id} \longrightarrow 3D_{obs}$$

Starting from the observed sky redshift space distribution of galaxy cluster, $Z_{obs}$, we generate a synthetic cluster distribution to account for unobserved clusters. This procedure allows us to reconstruct the whole-sky cluster distribution in the redshift space, $z_{id}$. An iterative reconstruction algorithm, similar to that proposed by Strauss & Davis (1988), is then applied to minimize the redshift space distortion, allowing us to recover the real-space distribution of the observed clusters $3D_{obs}$ and therefore also their peculiar velocities. The *intrinsic* reliability of the entire procedure has been tested using a simulated catalogue of clusters kindly provided to us by S.Borgani.

### 3.1 *z*-Space reconstruction scheme

To reconstruct the whole-sky redshift space cluster distribution, $z_{id}$, we adopted a phenomenological approach. The basic idea is to fill the artificially (due to Galactic absorption) low density regions of the surveyed volume with a synthetic cluster population having the same clustering properties as the distribution of the real clusters and then generate many Monte-Carlo realizations of this population. To this end we divide our procedure into two steps; one to reconstruct the cluster density field above $|b| = 20°$ and one to *fill* the ZoA.

#### 3.1.1 $|b| > 20°$

The first step is to recover the cluster distribution for $|b| > 20°$. We divided our volume into two regions: an inner sphere with a radius of R$_{in}$=200 $h^{-1}$ Mpc, which we consider as the region of reliable determination of the cluster density field, and an external region R$_{in} < r <$ 250 $h^{-1}$ Mpc.

In the inner part, in which $P(z) \simeq 1$, we used a Montecarlo rejection method to generate a population of synthetic clusters distributed according to the adopted $P(b)$ probability function and weighted them by $1/P(z)$ with the further constraint of being *spatially clustered around the real clusters* according to the observed spatial cluster–cluster correlation function (cf. Bahcall & Soneira 1983; Postman et al. 1992; Plionis, Valdarnini & Jing 1992). The following weights are then assigned both to real and mock clusters:

$$\tilde{w}_i(r) = \frac{\mathcal{W}_{rel}(r_i)}{P(z_i)}, \qquad (6)$$

Finally, the total number of synthetic clusters outside the ZoA is determined by requiring the overall number density of clusters to be:

$$\tilde{n} = \frac{1}{V} \sum_i \tilde{w}_i(z,b), \qquad (7)$$

where $\tilde{w}_i(z,b) = \mathcal{W}_{rel}(r_i) w_i(z,b)$. This requirement is fulfilled by imposing:

$$\sum_{real} \tilde{w}(z,b) = \sum_{real} \tilde{w}(r) + \sum_{synth} \tilde{w}(r), \qquad (8)$$

where the first two sums extend over real clusters and the last one over synthetic objects. Beyond 200 $h^{-1}$ Mpc the radial selection function, mainly for the ACO sample, falls below one roughly exponentially. Therefore weighting clusters that sparsely trace the density field with $1/P(z)$ can introduce non negligible shot noise errors. To avoid this problem and keeping in mind that we will limit our analysis within $R_{in}$ we reconstructed the cluster distribution in the external regions as following: the mock cluster distribution was Montecarlo generated to account for the $P(b)$ selection *with no spatial correlation with real clusters*; then the cluster redshift was determined according to $P(z)$ but, instead of weighting them by $1/P(z)$, we generated $1/P(z)$ synthetic clusters at the same $b$ and $z$ but with a randomly chosen $l$. Both real and synthetic clusters in this external region were simply weighted by $\mathcal{W}_{rel}(r)$ and their total number was determined by equating the local density with that of the central region.

### 3.1.2 Filling the ZoA

The second step is to fill the ZoA. The method we used is close to that proposed by S. Faber and used by Yahil et al. (1991). We divided the equatorial strip $-20° < b < +20°$ into 18 bins of 20° in longitude. In each bin we divided the distance range into bins of 2000 km/sec. The clusters were then sampled in the two northern and southern adjacent strips, defined so that their *joint* solid angle is equal to that of the ZoA. This requirement leads to the following limiting latitude for the strips:

$$|b_{lim}| = 90° - \cos^{-1}[2 \times \cos 70°] = 43.16° \ .$$

The number of synthetic clusters in each ZoA bin was then set equal to the sum of the clusters (real and synthetic) found in the two bins of the northern and southern, adjacent to the ZoA, strips; while their position, inside the ZoA bin, was randomly assigned. Finally, the real clusters in the range $13° < |b| < 20°$ are inserted and the total number of synthetic ZoA clusters is adjusted by subtracting a few at random until the average number density of real clusters is reached.

### 3.1.3 The $z_{id}$ Cluster Distribution

In what follows we will refer to the above 2-step $z$-reconstruction scheme as the *Randomized Standard Cloned Mask; [RSCM]*. At the end of the whole procedure nearly 50 % of the cluster population is composed by synthetic objects. A typical $z_{id}$ reconstruction is shown in panels (*a*) and (*c*) of Figure 1. Black dots represent real clusters while open dots are synthetic objects. The figure displays the projection of the reconstructed cluster distributions onto the supergalactic plane ($X_{sup}$-$Y_{sup}$ projection; panel *a*) and orthogonal to it ($X_{sup}$-$Z_{sup}$ projection; panel *c*). A simple visual inspection reveals that the synthetic clusters fill the ZoA and smoothly reproduce the clustering observed in the two nearby galactic-latitude strips. Outside the ZoA, as expected, synthetic objects are clustered with the true ones and their density decreases towards the galactic poles.

However, since each cluster has a different weight a more relevant visualization of the $z$-space cluster distribution is presented as smoothed density maps in panels (*b*) and (*d*) where we have overlayed a 20×20 grid (each cell being 2000 km/sec wide) on the cluster distribution of panels (*a*) and (*c*) and smoothed the projected distribution using a 2D–Gaussian kernel with $R_{sm} = 1$ cell, while weighting each cluster by $\tilde{w}_i$. What, in fact, we present in these figures is the 'mean' smoothed density field; an average over 10 random realizations of the $z$-space reconstruction procedure. Note that in panels (*b*) and (*d*) the cell units used have 0 and 20 corresponding to the Cartesian supergalactic coordinate -20000 and 20000 (in units of km/sec) respectively. It is interesting to note in panel (*d*) [$X_{sup}$-$Z_{sup}$ projection], the existence of a 'cross'-like structure with low-density cylindrical regions, passing through the cluster distribution and having a length of $\sim 20000$ km/sec. A more detailed discussion of the apparent structures will be presented in section 3.4.

### 3.1.4 Variants of RSCM

For the ZoA *filling* method we can use two variants of the previously described method. In the first variant, which we call *Randomized Heavy Cloned Mask [RHCM]*, we fill independently the northern and southern parts of the ZoA by applying the *[RSCM]* to the northern and southern adjacent galactic strips ($20° \leq |b| \leq 43.16°$) separately. With this mask the randomization applies to smaller bins and the cloning effect is expected to be enhanced.

In the second variant, which we call (*Randomized Thin Cloned Mask [RTCM]*), we identify the ZoA with the equatorial strip $|b| < 13°$. Then we consider the two adjacent strips defined so that their *individual* solid angles are equal to that of the ZoA. The *[RSCM]* is then applied by half counting the number of clusters in the adjacent bins. The purpose of this scheme is to account for the spatial clustering on larger volumes.

Finally, we also use an alternative scheme to reconstruct the $r > 200 \ h^{-1}$ Mpc region which is based in extending the reconstruction technique used for the $r < 200 \ h^{-1}$ volume to the external region. The ZoA *filling* method used is the standard one, described in the previous subsection. We call this scheme *Modified Standard Scheme [MSS]*.

### 3.2  3D–Space Reconstruction Scheme

Our 3D reconstruction scheme is based on the assumption that the peculiar velocities of clusters are caused by gravitational instability and that linear perturbation theory applies on the scales relevant to the cluster distribution. The distribution of clusters in redshift space differs from the true three–dimensional one by a nonlinear term in the redshift–distance relationship:

$$cz = H_\circ|\mathbf{r}| + [\mathbf{v}(\mathbf{r}) - \mathbf{v}(0)] \cdot \frac{\mathbf{r}}{|\mathbf{r}|}, \tag{9}$$

where $\mathbf{r}$ is the position of the generic cluster, $\mathbf{v}(\mathbf{r})$ its peculiar velocity, $\mathbf{v}(0)$ the peculiar velocity of the observer placed at the centre of the coordinate system and $H_\circ$ is the Hubble parameter.

#### 3.2.1  Linear Theory

In linear theory the peculiar velocities are proportional to the peculiar acceleration (cf. Peebles 1980)

$$\mathbf{v}(\mathbf{r}) = \frac{H_\circ f(\Omega_\circ)}{4\pi} \int \delta(\dot{\mathbf{r}}) \frac{\dot{\mathbf{r}} - \mathbf{r}}{|\dot{\mathbf{r}} - \mathbf{r}|^3} d^3 \dot{r} \ , \tag{10}$$

where $\delta(\mathbf{r}) = [\rho(\mathbf{r}) - \rho_b]/\rho_b$ is the mass density fluctuation about the mean $\rho_b$, $f(\Omega_\circ) \simeq \Omega_\circ^{0.6}$ and $\Omega_\circ$ is the cosmological density parameter at the present epoch. We assume that cluster density fluctuations are related to the mass fluctuations by a constant linear biasing factor:

$$\delta_c(\mathbf{r}) = b_c \delta(\mathbf{r}), \tag{11}$$

where $b_c$ is the bias parameter of clusters with respect to the mass. We can therefore replace $\delta(\mathbf{r})$ with $\delta_c(\mathbf{r})$ in eq.(10) provided that we substitute $f(\Omega_\circ)$ with $\Omega_\circ^{0.6}/b_c \ (\equiv \beta)$

In order to apply linear theory we need to smooth the discrete cluster density field on an appropriate scale, where non-linear effects could be important. Since we are using galaxy clusters to probe the density and the peculiar velocity fields, a natural smoothing scale, large enough for linear theory to be valid, would be the cluster correlation length (i.e. the distance $r_0$ at which the spatial cluster–cluster correlation function is unity). Recently Croft & Efstathiou (1994), analysing large N-body simulations, found that the velocity fields traced by galaxy clusters are highly non-linear below $\sim 10 \ h^{-1}$ Mpc. Therefore we allow the smoothing radius to vary in the range [10, 30] $h^{-1}$ Mpc. Although the cluster peculiar velocity field, the analysis of which will be presented in a forthcoming paper, does depend on the choice of the smoothing length, the dipole parameters depend only weakly on its choice, as we will see below.

The discrete cluster distribution was smoothed with a top hat window function which is equivalent to smooth the peculiar acceleration with:

$$W(|\dot{\mathbf{r}} - \mathbf{r}|) = \begin{cases} \frac{|\dot{\mathbf{r}} - \mathbf{r}|}{r_s^3} & \text{if } |\dot{\mathbf{r}} - \mathbf{r}| < r_s \\ 1 & \text{if } |\dot{\mathbf{r}} - \mathbf{r}| \geq r_s \end{cases} \tag{12}$$

where $r_s$ is the smoothing length.

Since $z_{id}$ is reconstructed up to $R_{max} = 250 h^{-1}$ Mpc, the integral in eq.(10) is converted into a sum over all the observed and synthetic clusters within $R_{max}$:

$$\mathbf{v}(\mathbf{r}) = \frac{\beta}{4\pi n_c} \tag{13}$$

$$\sum_i \left(\frac{M_i}{M_C}\right) W(|\dot{\mathbf{r}} - \mathbf{r}|) \tilde{w}(\dot{\mathbf{r}}_i) \frac{\dot{\mathbf{r}}_i - \mathbf{r}}{|\dot{\mathbf{r}}_i - \mathbf{r}|^3} + \beta \frac{\mathbf{r}}{3}.$$

The average cluster density $n_c$ is obtained by dividing the average cluster mass density $\rho_c$ with $M_C$ (mass of Coma cluster); $\rho_c$ is given by:

$$\rho_c = \frac{1}{V} \sum_i M_i \tilde{w}(r_i) \tag{14}$$

where $M_i$ is the mass of the generic cluster. As in PV91 and SVZ91 we assume that cluster masses are proportional to the the Abell–catalogue listed number of galaxies per cluster, $M_i \propto N_{A,i}$ (and $N_C = 106$); while the mass of the synthetic clusters was set equal to the average mass of the real clusters. The last term corrects for the net gravity of the homogeneous density background. The correction is necessary since the summation calculates the contribution from the density rather than from overdensity. Eq.(13) measures the true peculiar velocity only if density inhomogeneities outside the surveyed volume can be ignored.

As shown by Vittorio & Juskiewicz (1987) and Juskiewicz, Vittorio & Wyse (1990) in a multipole expansion of the external gravitational field the leading term is the dipole. Its effect is naturally removed if the peculiar velocities are evaluated in the LG frame because the dipole contribution is constant for all objects and does not affect the relative peculiar velocities. The only disadvantage of the LG frame is that the correction for the solar motion relative to its barycenter is uncertain by $\sim 50 - 100$ km/sec (Yahil et al. 1977) which, as we will see, is smaller than the errors on the predicted LG velocity. For this reason we used eq.(13) to evaluate the cluster peculiar velocity in the LG frame. The higher order terms in the multipole expansion of the external gravitational field fall off faster than the dipole with distance and are generally negligible for distances smaller than $R_{max}/2$ (Yahil et al. 1991). Possible systematic errors when computing peculiar velocities are therefore expected only at larger distances and should not appreciably affect the reconstruction reliability within 200 $h^{-1}$ Mpc.

### 3.2.2 The Reconstruction Algorithm

To reconstruct the 3D density field we used an iterative algorithm (which we call *Linear Iterative Reconstruction Algorithm, [LIRM]* hereafter), which is close to those described by Strauss & Davis (1988), Yahil et al. (1991) and Hudson (1993a). The idea is to start from the observed redshift space cluster distribution and attempt to recover the cluster true positions by iteratively computing their peculiar velocities according to eq.(10). This method is self consistent as long as the density field we are dealing with is linear, the clusters trace the mass and the fluctuations responsible for the peculiar motions of objects are within the volume sampled. Note also that in order to avoid problems related to the definition of distance and to the uncertainties on the Hubble constant, cluster distances are in km/sec units.

The steps of our iterative scheme are the following:

(i) All the clusters are initially placed at their observed distance, $r_z^{(0)} = cz$, with no peculiar velocity and with an arbitrary value for the input $\beta$ parameter. The index $^{(0)}$ refers to the zeroth iteration.
(ii) The weighting function $w(\mathbf{r}_i)$ is computed and its value is kept constant in the subsequent iterations (i.e. the selection function is not upgraded). This speeds up the algorithm while it also does not lead to appreciable errors, since $P(z) \approx 1$ within 20000 km/sec.
(iii) The average density $n_c$ within the sampled volume is computed and its value is updated.
(iv) The window function $W(|\mathbf{r}_i^{(k)} - \mathbf{r}^{(k)}|)$ is computed for each object.
(v) The peculiar velocity of the i-th cluster $\mathbf{v}^{(k)}(\mathbf{r}_i)$ is computed according to eq.(13) taking into account both, the clusters within $R_{max}$ and those that during the iterations were placed beyond the volume limits (we will refer to these as 'passive' objects).
(vi) Radial positions of 'active' clusters are updated. Their new positions are given by

$$r^{(k+1)} = cz - \left[\mathbf{v}(\mathbf{r})^{(k)} - \mathbf{v}(0)^{(k)}\right] \cdot \frac{\mathbf{r}^{(k)}}{|\mathbf{r}^{(k)}|}. \quad (15)$$

Positions of 'passive' objects are not updated.
(vii) For each real active cluster we compute the difference of its position between the last two iterations. If the difference is larger than 0.5% $r$, then we jump back to step (iii). The convergence is typically reached within 10 iterations.

Since the smoothing scheme treats clusters as spheres of radius $r_s$, clusters in the spherical shell $|\mathbf{r} - R_{max}| < r_s$ need to be weighted properly by computing how much of the sphere lies within $R_{max}$. But since we will estimate the dipole and the cluster velocity field only within 20000 km/sec and since we regard the cluster distribution beyond this radius as an improvement over a simple homogeneous distribution, we do not need to implement this correction.

Our iterative reconstruction algorithm can fail to recover the true 3D object position within high cluster density regions ('triple value regions') (cf. Yahil et al. 1991; Hudson 1993a). In our case, however, the objects are so sparsely distributed (which is the reason why the smoothing scale used is relatively large) that this problem is negligible. Obviously, the sparseness of the clusters introduces different problems, as we will discuss below.

## 3.3 Testing the reconstruction reliability

To test the *intrinsic* reliability of the whole reconstruction procedure and to optimize its performance we considered a catalogue of mock clusters extracted from a simulation performed by Borgani et al. (1994) in which the present time cluster positions were determined by the Zel'dovich approximation in a standard CDM model Universe. The mock clusters were identified as peaks above a density threshold which was chosen so that their space density matches that of the observed clusters ($\sim 2 \times 10^{-3} h^3$ Mpc$^{-3}$). We considered a spherical volume centered around a LG–like observer (see Borgani et al. 1994). Each cluster peculiar velocity was determined according to linear theory after having computed the gravitational acceleration generated by all the clusters within the sampled volume. This ideal cluster distribution was then degraded by Montecarlo rejecting clusters according to the probability distributions $P(z)$ and $P(b)$ and by devoiding the ZoA. Finally, an artificial Gaussian distributed noise was added to the cluster velocity to mimic experimental errors in their measured redshift. According to observational indications we used a variance of 200 km/sec (Strauss, Cen & Ostriker 1994). Finally we applied the *RTCM* together with the *LIRM* to recover the original object distribution and velocities.

Three different indicators were used to assess the *intrinsic* reliability of the reconstruction procedure:

- The difference between reconstructed and true peculiar acceleration acting on the observer.
- The volume averaged discrepancy between the true and the reconstructed position: $\langle \epsilon_i \rangle = \langle |\mathbf{r}_{true,i} - \mathbf{r}_{rec.,i}|/|\mathbf{r}_{true,i}| \rangle$
- The spatial cluster–cluster correlation function.

The reconstructed peculiar acceleration vector is very close to the original one. The per cent discrepancy between the amplitudes of the true and the reconstructed peculiar velocities, based on 10 different reconstructions, was $3 \pm 6\%$. The two vectors were misaligned by $2° \pm 5°$. In Figure 2 we present the distribution of the $\epsilon_i$ values which peaks at $\lesssim 1\%$ and has a mean value of $\langle \epsilon_i \rangle \simeq 1.6\%$ (it is skewed towards large errors); while the probability of having $\epsilon_i > 3\%$ is quite low. We found, as expected, that $\epsilon_i$ depends on the initial displacement of the $i$-th cluster with respect to its true position $|\mathbf{r}_{true,i} - \mathbf{r}^{(0)}|$. The reliability of the method greatly improves if clusters are initially placed close to their real positions, which would be possible if reliable redshift independent distance measurements were available for some nearby clusters. Finally, the spatial two point correlation function of the reconstructed cluster distribution turns out to be indistinguishable from the true one.

We stress that these tests are aimed at assessing the *intrinsic* reliability of the reconstruction method; cluster velocities are computed according to linear theory and no allowance for contribution from external gravitational fields is made (a set-up that needs not be true in the real Universe). To get an estimate of the actual uncertainties in the whole procedure we need to account for several other effects, which will be quantified in Section 5. The *intrinsic* uncertainty will be assumed to be independent from the other source of errors and will be added in quadrature to compute the total uncertainty on the peculiar velocities.

### 3.4 Smoothed cluster density maps in $z$- and 3D–space

In Figure 3a and b we plot the reconstructed whole-sky smoothed density field for a slice of 8000 km/sec wide, centered on the supergalactic plane, in $z$-space and in 3D space, respectively. Note that the same Gaussian smoothing and cell units are used as in figures 1b and 1d. A different representation of the cluster density in this slice is also presented in Plate 1, where we plot a 3D surface-plot visualization of the density field (similar to that presented by Dekel 1994), again in both $z$-space (upper panel) and 3D (lower panel). As also outlined by Scaramella (1994) who derived and discussed similar maps but only in redshift space, the sparseness of the cluster distribution and the heavy smoothing could miss some features of the cosmic density field. Nevertheless, our reconstruction technique should reveal the main large-scale features of the 3D cosmic density field.

From these plots it is evident that eliminating the distortions due to peculiar velocities suppresses significantly the amplitude of the density peaks in 3D with respect to that in the $z$-space. This effect is particularly evident in the Virgo-Hydra-Centaurus (or else Great Attractor) region [centered at $(X,Y) = (8.5,11)$], in the Leo region [centered at $(X,Y) = (11,14)$] and around the Ursa-Major supercluster [centered at $(X,Y)=(15,18)$]. Interestingly, in the Perseus-Pegasus region [centered at $(X,Y)=(13,10)$] the opposite is true, ie., the 3D density amplitude is slightly higher than in the $z$-space case. This could be easily understood if there exists a coherent gradient of negative peculiar velocities, increasing in amplitude towards the near side of this structure (towards smaller X's, ie., towards the LG).

Also, the shape of the density peaks, in the 3D case, appears to be slightly elongated along the line of sight. This effect, which is mostly prominent in the Great Attractor and in the Shapley concentration [centered at $(X,Y)=(3.5,14)$], arises from the fact that in the linear regime the infall peculiar velocities within high density regions are coherent and tend to twist the isodensity contours along the line of sight.

In figure 4 we present the 3D cluster density field for the four slices, each of 8000 km/sec width, above and below the supergalactic plane. Panel ($a$) and ($b$) correspond to the $4000 < Z < 12000$ km/sec and $12000 < Z < 20000$ km/sec slices respectively, while panel ($c$) and ($d$) represent the corresponding slices but for negative Z's. Our intention is not to present a detailed analysis of the structures evident in our contour plots (for such a task see Tully 1987 and Tully et al. 1992). However, we would like to emphasize the fact that, most structures found in the supergalactic plane extend not only to the first 8000 km/sec wide slice, above or below the supergalactic plane, but also to the second such slice. For example the overdensity corresponding to the Shapley concentration extends to both negative Z slices. Also the Great Attractor region is connected to the Lepus region [centered at $(X,Y)=(8.5,10)$ in panel $c$] which seems to extend even further to the next slice (panel $d$) while a similar behaviour is found also for Perseus-Pegasus region. From panels ($a$) and ($b$) we see that the same is true for the Ursa Major supercluster, the Grus-Indus region [centered at $(X,Y)=(3,2.5)$ in the supergalactic plane slice], the Leo region which seems to be connected with Hercules [centered at $(X,Y)=(10,14)$ in the ($a$) and ($b$) slices] as well as the Pegasus-Pisces region [centered at $(X,Y)=(17.5,4)$] which is evident in the supergalactic plane and the two positive Z slices.

## 3.5 Redshift space distortions of the dipole vector

One of the major scopes of this work is to compare the cluster dipole as measured in $z$-space with the reconstructed one in 3D–space. A simple eye inspection of figures 3 a,b and Plate 1 revealed that the peaks of the true density field have larger amplitudes when observed in redshift space. However, to see how these distortions affect the estimate of the cluster dipole require a quantitative discussion. From a theoretical point of view the problem has already been addressed by Kaiser (1987). He showed that gravitationally driven peculiar velocities can have a large effect on the peculiar acceleration vector of an object. If $g_r$ is the amplitude of the true peculiar acceleration and $g_z$ is the $z$-space peculiar acceleration, then the relationship between $g_r$ and $g_z$ depends both on the selection function $\phi$ and on the on the power spectrum considered. If one considers each power mode separately then, in the limit $kR_{max} \gg 1$ ($R_{max}$ is the depth of the survey and $k$ is the wavenumber), Kaiser has shown that

$$g_z - g_r = g_r \frac{\Omega_\circ^{0.6}}{3}[1 + E(k)] , \qquad (16)$$

where

$$E(k) \simeq \log[(kR_{max})^2 \phi(R_{max})/\phi(k^{-1})] . \qquad (17)$$

Clearly the sign of $E(k)$ depends on the selection function and on the sample's depth. Although a precise $E(k)$ evaluation require a good determination of the selection function (that, for the clusters case, also depends on the relative weight), we can be confident that in our case $E(k) > 0$. In fact for the Abell and ACO cluster sample $\phi(r) \simeq 1$ within the sample (the present analysis is limited to a depth of $R_{max} = 200\ h^{-1}$ Mpc). In the linear regime the final value of $g_z - g_r$ can be obtained by combining each mode separately (for large $k$ the nonlinear evolution couple the different modes, and this approximation does not hold). Obviously to properly combine the different modes requires the knowledge of the density fluctuation power spectrum. However, unless one considers cosmological models with very large power on small scales, which are in conflict with current observations, we expect that for our cluster sample:

$$g_z - g_r > 0 \qquad (18)$$

## 4 THE CLUSTER DIPOLE

The dipole vector, for an observer placed at the center of the coordinate system, is defined as

$$\mathbf{D} = \frac{3}{4\pi} \int \rho_c(\mathbf{r}) \frac{\mathbf{r}}{|\mathbf{r}|^3} d\mathbf{r}, \qquad (19)$$

which, in linear theory is related to the peculiar velocity by

$$\mathbf{v} = \beta \frac{H_\circ \mathbf{D}}{3\langle \rho_c \rangle} . \qquad (20)$$

For our discrete cluster field density field, smoothed with a top hat window, the dipole becomes:

$$\mathbf{D} = \frac{3M_C}{4\pi} \sum_i \left(\frac{M_i}{M_C}\right) w(r_i) W(r_i) \frac{\mathbf{r}_i}{r_i^3} \qquad (21)$$

The monopole of the smoothed field is defined as:

$$M = \frac{M_C}{4\pi} \sum_i \left(\frac{M_i}{M_C}\right) w(r_i) W(r_i) \frac{1}{r_i^2} \qquad (22)$$

Since the value of $\mathbf{D}/M$ has a smaller scatter than that of $\mathbf{D}$ (see PV91) it can be used to compute the LG peculiar velocity, assuming that the dipole has converged to its asymptotic value within the volume sampled. In this case the relationship between the peculiar velocity and $\mathbf{D}/M$ is:

$$\mathbf{v}(r) = \beta \frac{\mathbf{D}(r)}{3} \frac{H_\circ R_{conv}}{M(R_{conv})} , \qquad (23)$$

where $R_{conv}$ is the final dipole's convergence depth. However, based on Montecarlo experiments, we found that a better estimate of the peculiar velocity of the observer is obtained directly from eq.(13). The average $1\sigma$ scatter between $\mathbf{v}$, computed from eq.(23), and the true one is $\sim 12\%$ if 850 'clusters' are used, a number comparable to that of our cluster (true and synthetic) sample. The discrepancy originates from the fact that the monopole is a good estimator of the average density only in the limit of continuous density field. Using 8500 objects, for example, reduces the scatter to 7%. This effect could introduce a further uncertainty in the LG peculiar velocity computed via eq.(23).

Computing the dipole and the monopole separately is nevertheless important for assessing the effective convergence of the peculiar velocity vector since a necessary condition for final convergence is that while the dipole converges, the cumulative monopole continues to grow linearly with distance. This condition, however, is not sufficient to guarantee the final convergence of the dipole, which could still increase after having reached a *plateau* (cf. Plionis, Coles & Catelan 1993).

### 4.1 Possible Systematic Effects

The reconstructed cluster distribution could depend on a number of free parameters that are only partially determined by observational constraints and theoretical requirements. We systematically explored the influence of the parameter choice by allowing each free parameter to vary within an experimentally plausible range. In practice the reconstructed cluster dipole, and therefore the predicted LG peculiar velocity, can be regarded as a function of many variables (the free parameters) that are listed in table 1. Boldface quantities refer to what we call the *standard case*, defined by the following choice of parameters: $(\alpha, \alpha) = (0.3, 0.2)$, $r_s = 20\ h^{-1}$

Mpc, $q_0 = 0.5$, $\beta = 0.25$, direct homogenization scheme (eq.5) with the 200 $h^{-1}$ Mpc sphere being divided into 10 equal volume shells. For each of the 48 possible combinations of the free parameters, to which we refer to as "models", we performed 10 independent reconstructions and we computed the corresponding 3D dipole. To explore the influence of each variable on the predicted LG dipole separately we adopted the following procedure:

- The parameter of interest is selected and models are divided according to the value chosen for that parameter (e.g. $q_0 = 0.2$ or 0.5). Models having the same parameter values are grouped in the same category.
- The average dipole at 170 $h^{-1}$ Mpc is computed for all the models in the same category and the same procedure is repeated for all different categories.
- The different average dipoles are compared. The differences in amplitude and the misalignment angles between different average dipoles are listed in table 2.

It is evident from table 2 that the only appreciable systematic effect comes from the choice of the $\beta$ parameter, as we will discuss below. Varying the other experimental parameters (i.e. the galactic absorption coefficients, the smoothing radius and the homogenization procedure) generates systematic errors much smaller than the intrinsic uncertainties of the reconstruction.

### 4.1.1 The effect of varying the $\beta$ parameter

To appreciate the influence of the $\beta$ parameter on the reconstruction process we have to disentangle two different effects: the role of the deceleration parameter, $q_0$, that enters through eq.(1) in defining the outer limit of the sample, and that of the $\beta$ factor, whose arbitrary initial value has to be specified for the *LIRM* to be applied. The first effect turned out to be much smaller than the second one.

The effect of varying the $\beta$ parameter in the *LIRM* was explored by performing the reconstruction with the same $q_0$ value but using two different $\beta$ values (0.25 and 0.114). Although these values are different by a factor $\sim 2$, the asymptotic amplitudes of the reconstructed dipole are very similar, differing only by 8%, being smaller for the standard $\beta = 0.25$ case, while the direction does not change appreciably. The complete analysis was performed by exploring two different class of models having ($\beta = 0.25$, $q_0 = 0.5$) and ($\beta = 0.114$, $q_0 = 0.2$), respectively, confirmed the above results (see table 2). It is therefore clear that there is a small, although systematic, effect related to the value of the $\beta$ parameter which enters as an input in the reconstruction algorithm.

It is interesting to compare this result to that obtained by Strauss et al. (1992, [ hereafter SYDHF92]). The asymptotic amplitude of their IRAS galaxy dipole, on scales larger than $\sim 100$ $h^{-1}$ Mpc, strongly depends on the input $\beta$ value. The reason for this behaviour was ascribed to redshift space distortions, to which they refer as the 'Kaiser effect', that in their case is mainly due to the LG motion with respect to the distribution of distant galaxies. In this simple situation the problem is similar to that of a 'rocket–born' observer who resides in a uniform Universe at rest with respect to the CMB on large-scales and whose peculiar acceleration originates locally (Kaiser & Lahav 1988). In this idealized case the redshift space distortions appreciably affect the measured peculiar acceleration only if the selection function $\phi$ falls below one, well within the sample's depth. This is the case for the IRAS galaxy sample studied by SYDHF92, where $\phi \ll 1$ well within 100 $h^{-1}$ Mpc. For our sample, however, $P(z)$ is close to unity within 200 $h^{-1}$ Mpc so that the 'Kaiser Effect' is much less important, as demonstrated by the stability of our dipole against the variations of $\beta$; especially at large depths where the sensitivity to the input $\beta$-parameter would be manifested by variations of the dipole amplitude shape as a function of $\beta$ (cf. Fig.6 of SYDHF92).

## 4.2 Error Estimate

To obtain a reliable error estimate of the reconstructed dipole we need to account for several different sources of errors. We estimate the *intrinsic* error, $\sigma_i$, in the reconstruction algorithm from the 10 independent reconstructions performed for each model explored. Other sources of errors, which will be described below, are the observational error $\sigma_o$, the 'shot–noise' error $\sigma_{sn}$ and the zero point uncertainty $\sigma_{zp}$. In what follows we will assume that all these errors are independent so that the total uncertainty $\sigma_T$ will be simply estimated by adding the errors in quadrature.

### 4.2.1 Observational error

We define as observational error the uncertainty derived from the different free parameter choice, listed in table 1, whose separate influence on the reconstructed dipole was explored in the previous subsection. The reconstructed dipole can be therefore regarded as a function of the smoothing length, the homogenization scheme and its binning, parameters that we allowed to vary in a plausible range of values. The galactic absorption set and the $\beta$ factor, instead, are considered fixed by observations to $(0.2, 0.2)$ and to $\beta = 0.25$, respectively. We chose $\beta = 0.25$ since, as we will see in section 5, this value is closer to that one derived from the comparison with the CMB dipole. We found that varying these parameters does not produce any systematic effect in the dipole determination and thus we can consider them as Gaussian variables and the dipole as a multivariate

Gaussian function of the explored parameters. Following this argument we compute $\sigma_0$ as the variance of the dipole obtained by averaging over all models having $(\alpha, \alpha) = (0.3, 0.2)$ and $\beta = 0.25$ (our standard case). The observational error at 170 $h^{-1}$ Mpc produces an uncertainty of $\sim \pm 65 \, \beta^{-1}$ km/sec in the dipole amplitude and of $\pm 4°$ and $\pm 2°$ along $b$ and $l$ directions respectively.

*4.2.2 Shot-Noise error*

The usual shot–noise estimate assumes that luminous objects have been drawn by a Poisson process from the underlying density field. However the reconstruction procedure relies on the hypothesis that clusters of galaxies are *biased* tracers the mass. A more meaningful and self–consistent estimate of the shot noise error has been proposed by SYDHF92 that accounts for two different effects:

- The Poissonian error due to the fact that luminous objects become random tracers of the density field when the selection function falls below unity.
- An uncertainty due to the fact that cluster masses are randomly taken from an underlying mass distribution. In our case we estimated the mass of real clusters from their Abell listed number of galaxies while for synthetic clusters, which constitute $\sim 50\%$ of the whole cluster population, the mass was set equal to the average real cluster mass.

Using our formalism, the SYDFF92 shot noise computed for the $\mathcal{W}_{rel} = 1$ case, reduces to:

$$\sigma_{sn}^2 = \left(\frac{\beta}{4\pi n_c}\right)^2 \sum_i^{N_s} \frac{W(r_i)^2}{r_i^4} K\left[\tilde{w}(r_i) - 1\right], \quad (24)$$

where the sum extends only over the $N_s$ synthetic clusters, $n_c$ is the average cluster density defined from eq.(14) and $K = \langle \mu_i^2 \rangle - 1$, with $\mu_i$ being the mass of the real clusters in units of average mass. From the real clusters mass variance we have K=0.25. To compute the influence of the shot noise on the amplitude and the direction of the dipole we will assume that each component of the dipole is a Gaussian with zero mean and $\sigma_{sn,x} = \sigma_{sn,y} = \sigma_{sn,z}$. The 1-dimensional shot noise error at 170 $h^{-1}$ Mpc reflects in an $\sim \pm 60 \, \beta^{-1}$ km/sec amplitude and a $\sim \pm 4°$ directional uncertainty.

*4.2.3 Zero point error*

The gravitational influence on the LG motion from nearby clusters, such as the Virgo cluster which was not included in the Abell/ACO catalogue due to its proximity and therefore of its large angular dimension, are not negligible. Neglecting them in the cluster-dipole analysis, causes a zero point offset on the predicted LG peculiar acceleration. A possible way to overcome this problem is to merge the cluster dipole with that of some galaxy sample that trace more densely the local mass distribution (Scaramella Vettolani & Zamorani 1994, [SVZ94 hereafter]). With this approach, however, the offset can be evaluated only assuming *a priori* a value for the relative bias of the cluster and galaxy populations, whose value is very uncertain (for a complementary approach see Plionis 1995 *in preparation*).

Here we adopt a simpler phenomenological approach by identifying the zero point offset with a 200 km/sec LG Virgocentric infall to which we attach a $\pm 100$ km/sec uncertainty. This results in an uncertainty of $^{+0.5}_{-1}$ degrees and $^{+8}_{-11}$ degrees in the dipole direction along $l$ and $b$, respectively. Analogously, for the dipole's amplitude we obtain an uncertainty of $^{+55}_{-40} \, \beta^{-1}$ km/sec.

### 4.3 The 3D Cluster Dipole

After reconstructing the cluster distribution with the *RSCM* and the *LIRM*, we computed the cluster dipole both in redshift and real–space. Figure 5a shows the amplitude of the peculiar velocity for the standard model, computed using eq.(13), as a function of distance. Open dots refer to the $z$-space dipole while filled symbols represent the reconstructed 3D cluster dipole. We also plotted, for reference, the dipole computed without any Abell/ACO homogenizing scheme (i.e. $\mathcal{W}_{rel} = 1$) in $z$-space (continuous line). Errorbars represent 1 $\sigma$ total errors.

Our main result is that the asymptotic value of the 3D dipole is significantly less than the $z$-space dipole, as predicted by eq.(18). Removing redshift space distortions erases the artificial redshift space clustering, leading to a smaller, by $\sim 23\%$, asymptotic dipole amplitude. Although the amplitude of the 3D dipole is significantly less than what previously found in cluster-dipole $z$-space studies (PV91; SVZ91), the qualitative dipole behaviour is similar in $z$-space and 3D–space: a "bump" of the velocity amplitude around $50 h^{-1}$ Mpc followed by a decrease due to the competing pull of the Great Attractor with the Perseus–Pegasus regions, a secondary increase after a *plateau* and an asymptotic convergence beyond $\sim 170 \, h^{-1}$ Mpc once that the Shapley concentration has entered into the sampled volume. The mass distribution beyond 60 $h^{-1}$ Mpc is responsible for $\sim 30\%$ of the total LG peculiar acceleration. As already noticed from the qualitative analysis of the density field maps, the effect of eliminating the apparent $z$-space clustering is particularly important in local high density regions, such as the GA area, which is clearly responsible for the significant decrease of the $\sim 50 \, h^{-1}$ Mpc bump in the 3D cluster dipole case. In Figure 5a we also plot, in arbitrary units, the 3D space monopole term, which grows linearly out to the depths sampled which indicates that the dipole convergence at $\sim 170 \, h^{-1}$ Mpc is not an artifact of ill sampling.

Figure 5b displays the cumulative direction of the LG velocity. The starred symbol indicates the CMB dipole apex, while the skeletal symbol represents the CMB direction after correcting the LG for a Virgocentric infall of 200 km/sec. Errorbars represent the 1 $\sigma$ total errors. The relatively large error ($\sim 10°$) in the dipole direction, along $b$, is due to the uncertainty related to the amplitude of the Virgocentric infall. The reconstructed 3D dipole points $\sim 10°$ away from the CMB apex when the Virgo infall is taken into account. No significant differences between the 3D and $z$-space dipole directions was found.

#### 4.3.1 The effect of varying the Mask model

It is interesting to investigate the influence of the Mask model, adopted to reconstruct the whole sky $z$-space cluster distribution, on the reconstructed dipole. Here we compare the $RSCM$ reconstructed standard dipole with the two obtained by using the same set of standard parameters but adopting the $RHCM$ and $RTCM$ mask models. In both cases there is a small increase of the dipole's amplitude, by $\sim 3\% - 4\%$, and the direction points $\sim 4 \pm 4°$ away from the original one (uncertainties refer to the intrinsic errors). We conclude that no significant dependance was found on the Mask choice, which is to be expected since all of them are based on a cloning scheme.

A more relevant comparison is probably between two conceptually different methods. For example between our $RSCM$ and the spherical harmonic reconstruction method (cf. Lahav 1987; Yahil et al 1986; Plionis 1988 and PV91 for use of this method in the cluster dipole context). In Figure 6 we compare our standard $RSCM$ $z$-reconstructed dipole with that derived using the PV91 method for the same sample and the same set of parameters. In the insert we plot the relative velocity fluctuations, defined as $2 \times (v_{RSCM} - v_{SH})/(v_{RSCM} + v_{SH})$. It is evident that both methods give identical results, except in the inner volume ($\lesssim 50\ h^{-1}$ Mpc) where the number of clusters is extremely small and therefore shot noise effects are large. This is quite extraordinary given the completely different methods used.

Finally, we explored the effect of using the $MSS$ with which the reconstruction scheme adopted in the inner region ($\lesssim 200\ h^{-1}$ Mpc) is extended up to the limiting sample depth (250 $h^{-1}$ Mpc). With the $MSS$ the 3D dipole differs from the standard one only beyond 200 $h^{-1}$ Mpc, where the amplitude monotonically increases (by $\sim 8\%$) and the direction systematically drifts towards lower galactic latitudes. As we discussed in section (3.1), using the $MSS$ increases the shot noise errors in a region where the selection function is significantly smaller than 1. The shot noise errors are probably responsible for the drift in the dipole direction, while they also enhance the 'Kaiser Effect' which causes the observed amplitude increase.

## 5 COSMOLOGY FROM THE CLUSTER DIPOLE

### 5.1 Constraints on the value of $\beta$

From the measured asymptotic value of the cluster dipole it is possible to constrain the value of $\beta$. An estimate of this parameter can be obtained by simply assuming that the cluster 3D dipole has converged at $\sim 170\ h^{-1}$ Mpc, where it flattens, in accordance with SVZ91 and PV91. Using eq.(13) to compare the dipole amplitude with the LG motion, corrected for the Virgocentric infall leads to a value of the $\beta$-parameter:

$$\beta = 0.21^{+0.07}_{-0.04}$$

If we erroneously did not account for the Virgo infall we would obtain $\beta = 0.25^{+0.08}_{-0.05}$. Errors come from total dipole uncertainties, as discussed in section 4.

Although the dipole convergence at $\sim 170\ h^{-1}$ Mpc is real, since at that distance the cluster monopole is still linearly increasing, it is impossible to prove that the dipole has converged to its final value. The only way out is to evaluate the amplitude of the external contribution to the observed dipole using statistical methods.

Juskiewicz, Vittorio & Wyse (1990) computed the probability distribution of the peculiar velocity of a generic observer, constrained to be moving with a velocity of 620 km/sec with respect to the CMB, as a function of the sampled volume. From this distribution they computed the expectation value for the estimated LG velocity and its variance (their eq. 8a and 8b). The expectation value at 170 $h^{-1}$ Mpc weakly depends on the cosmological model adopted. We explored a number of possible models ranging from the Standard CDM one to the phenomenological model proposed by Branchini et al. (1994) that resembles a tilted CDM with $n = 0.7$ and to the Peacock & Nicholson (1994) spectrum with more power on large scale (similar to a standard CHDM model). In all these cases the expectation value for the LG velocity, at 170 $h^{-1}$ Mpc coincides with the true one. Also the variance slightly depends on the cosmological model, being larger in correspondence to the Peacock & Nicholson (1994) spectrum ($\sim 600\ \beta^{-1}$ km/sec that, to be conservative, will be used as the typical cosmological variance). A probably better, although exaggerated, estimate of the $\beta$ uncertainty can be therefore obtained by adding in quadrature the cosmological variance to the total error. The resulting $\beta$ value is thus constrained into the range

$$\beta = 0.21^{+0.09}_{-0.11}\ .$$

The 1 $\sigma$ range is wide enough to overlap with previous estimate of $\beta$ based on the $z$-space cluster distribution

(PV91, SVZ91, SVZ94). Assuming $\Omega_o = 1$ we obtain a value for the cluster bias parameter

$$b_c = 4.8^{+3.5}_{-1.1}.$$

According to the linear biasing prescriptions if $b_A$ and $b_B$ are the biasing factors of objects $A$ and $B$ with respect to the matter with $b_A > b_B$, the following relation holds: $b_{AB} \sim b_A/b_B$, where $b_{AB}$ is the biasing factor of objects $A$ with respect to objects $B$. If the biasing factor of Abell/ACO clusters with respect to IRAS galaxies is $b_{cI} \sim 4.5$ (cf. Peacock & Dodds 1994; Jing & Valdarnini 1993) we have that the above $b_c$ value is consistent with a bias parameter of IRAS galaxies with respect to the mass of $b_I \sim 1.1^{+0.7}_{-0.3}$.

## 5.2 Constraining Dark Matter Models

The observed cumulative acceleration acting on the LG can be used to discriminate between different cosmological models. Useful constraints can be imposed by requiring a cosmological model to reproduce the observed cumulative dipole. As outlined by SVZ94, the dipole increase in the range beyond 80 $h^{-1}$ Mpc is a characteristic feature which can be used as a cosmological test. To quantify the constraint we consider the ratio of peculiar velocity $\chi_{(R_1,R_2)}$ induced on an observer placed on the origin from fluctuations within a sphere of radius $R_1$ to that from within a sphere of radius $R_2 > R_1$. Juskiewicz, Vittorio & Wyse (1990) computed the probability distribution for this ratio as a function of the cosmological model through the power spectrum $P(k)$. Here we computed the probability density distribution for ratio $\chi_{(R_1,R_2)}$ of the peculiar velocities generated within the two spheres of $R_1 = 100$ and $R_2 = 160$ $h^{-1}$ Mpc. The probability distributions computed for the three cosmological models previously used, are displayed in Figure 7. The continuous line represents the experimental values from the 3D standard dipole in Figure 6 assuming a Gaussian distributed total errors. Short dashed line represents the CDM standard model, long dashed line refers to the Branchini, Guzzo & Valdarnini (1994) phenomenological model while the Peacock and Nicholson (1991) model is displayed as a dot–dashed line. As already noticed by SVZ94, models with a large scale of coherence are preferred by the observed cluster dipole even when redshift space distortions are removed. A detailed analysis, using large Zel'dovich simulations of the cluster distribution, on the constraints put by the cluster dipole on the different Dark Matter models will be presented in Tini Brunozzi et al. (1995).

## 6 CONCLUSIONS

In this work we described a self consistent method to reconstruct the 3D distribution and peculiar velocities of galaxy clusters from their observed positions in the redshift space. Our approach is valid within the framework of gravitational instability and relies on the assumption of linear theory and linear biasing. The technique presented aims at recovering, in a statistical sense, the distribution of unobserved clusters and corrects the apparent cluster positions for peculiar velocity driven redshift space distortions. Tests performed using synthetic cluster distributions, obtained from numerical experiments, showed that the intrinsic reliability of this technique allows the true cluster positions to be recovered with an average error of $\sim 1.6$ %, much better than the best available experimental determination of cluster distances ($\gtrsim 10\%$). However, the actual reconstruction uncertainty is larger since we adopted a phenomenological method, to reconstruct the whole sky $z$-space distribution of clusters, characterized by several parameters with associated observational errors. As necessary in any phenomenological approach, we tested the stability of the final results on the various experimental parameters and found them to be surprisingly robust.

Our main result is that redshift space distortions cause an artificial increase of the cluster dipole amplitude by $\sim 23\%$, while they do not affect appreciably the dipole direction which points $\sim 10°$ away from the CMB apex. The cumulative 3D dipole qualitative matches the behaviour of the $z$-space one: there is a bump at $\sim 50$ $h^{-1}$ Mpc and, after a *plateau*, it rises by $\sim 30\%$ and stabilizes beyond 170 $h^{-1}$ Mpc where it reaches a real, although not necessary final, convergence. The reduction of the 3D dipole amplitude with respect to that of the $z$-space one arises mainly from the removal of the strong apparent clustering in the Great Attractor and Shapley regions and its enhancement in the Perseus-Pegasus region, as can be clearly seen in the smoothed density maps of Figure 3a,b and Plate 1. These results are robust in the sense that a search for the presence of systematic biases gave negative results. In practice we found that the freedom of modelling the galactic absorption, the radial selection function, the ZoA filling scheme, the homogenization and the smoothing procedures cause systematic errors much smaller than the intrinsic error in the reconstruction procedure. Interestingly, since the selection function of our cluster sample is $\sim 1$ up to $\sim 190$ $h^{-1}$ Mpc, our 3D dipole determination is only marginally affected by the 'Kaiser effect' which has been found to be a serious problem in the 3D galaxy dipole estimations.

From the asymptotic dipole amplitude it is possible, in principle, to determine the $\beta$ parameter. The intrinsic cosmological variance, however, reflects in a large uncertainty whose magnitude depends on the unknown cosmological background. In the conservative and widely accepted hypothesis that the cosmic density field is characterized by a large correlation length, we estimated $\beta = 0.21^{+0.09}_{-0.11}$. This value is larger than previously found based on the analysis of the $z$-space cluster distribution, the reason being that we have accounted for redshift space distortions. The large $1\sigma$ uncertainty, however, does not allow us to set stringent limits on the value of $\beta$. Better constraints will be presented in a forthcoming paper in which we will analyze the reconstructed cluster peculiar velocities and compare them to those determined observationally.

### Acknowledgements

We would like to thank Prof. Dennis W. Sciama for suggesting to us a problem out of which the idea for this work was born. We also thank Dr. Stefano Borgani for giving us access to his Zel'dovich cluster simulations. MP acknowledges receipt of an EEC *Human Capital and Mobility Fellowship*.

**Table Captions**

**Table 1:** Experimental parameters. For each parameter the values adopted are shown. Each model explored is characterized by a different set of parameters. Boldface quantities represent the 'Standard Model' set.

**Table 2:** Differences between the different average dipoles. First column displays the parameters defining the two compared categories. Column 2 shows the per cent difference of the dipole amplitude at 170 $h^{-1}$ Mpc together with the relative standard deviation. Columns 3 and 4 contain the average misalignment and its standard deviation at 170 $h^{-1}$ Mpc along $l$ and $b$, respectively.

**Figure Captions**

**Figure 1:**

(a) and (c): The projected whole-sky $z$-space cluster distribution in Cartesian supergalactic coordinates. Filled symbols represent real Abell/ACO clusters while open symbols the synthetic objects. Short dashed lines delineate the $|b| \leq 20°$ region. (a) $X_{sup}$-$Y_{sup}$ projection and (c) $X_{sup}$-$Z_{sup}$ projection.

(b) and (d): The projected smoothed whole-sky $z$-space cluster density field (of panels (a) and (c)). A 2–D Gaussian kernel was used with $R_{sm} = 2000$ km/sec. (b) $X_{sup}$-$Y_{sup}$ projection and (d) $X_{sup}$-$Z_{sup}$ projection.

Note that cell units are used (see text for definitions).

**Figure 2:** The distribution of relative percentage differences, $\epsilon_i$, between the true and reconstructed positions of simulated clusters (see text for details) which were used to assess the reliability of our reconstruction method.

**Figure 3:** The projection onto the supergalactic plane of the smoothed cluster density field with $-4000 < Z_{sup} < 4000$ km/sec. (a) $z$-space (b) reconstructed 3D-space. Same smoothing procedure used as in figure 2.

**Figure 4:** $X_{sup}$-$Y_{sup}$ projections of the smoothed 3D-space cluster density field for 4 slices in $Z_{sup}$. (a) $4000 < Z_{sup} < 12000$ km/sec. (b) $12000 < Z_{sup} < 20000$ km/sec. (c) $-12000 < Z_{sup} < -4000$ km/sec. (d) $-20000 < Z_{sup} < -12000$ km/sec.

**Figure 5:** Cluster dipole.

(a) Dipole amplitude as a function of distance with open symbols representing the reconstructed whole-sky $z$-space dipole and filled ones the reconstructed 3D dipole. The continuous thin line refers to the $z$-space dipole without any Abell/ACO homogenizing scheme (i.e. $\mathcal{W}_{rel} = 1$). In arbitrary units, we plot also the 3D-space monopole (small filled squares).

**(b)** Dipole misalignment angle. The starred symbol indicates the CMB dipole apex, while the skeletal symbol represents the same but after correcting for a Virgocentric infall component of the LG motion, with an amplitude of 200 km/sec.

Errorbars are 1 $\sigma$ total uncertainty (see text for details).

**Figure 6:** Comparison of the $z$-space dipole derived by using our standard $RSCM$ method to reconstruct the whole-sky cluster distribution and that derived using the spherical harmonic method (cf. PV91) for the same sample and the same set of parameters. In the insert we plot the relative velocity fluctuations between the two determinations.

**Figure 7:** The probability density distribution for the dipole amplitude ratio between the two spheres with radii $R_1 = 100$ and $R_2 = 160$ $h^{-1}$ Mpc. The continuous line represents the experimental values from the 3D standard dipole of Figure 7. Short dashed line represents the CDM model, long dashed line refers to the Branchini, Guzzo & Valdarnini (1994) phenomenological model while the Peacock and Nicholson (1991) model is displayed as a dot–dashed line.

**PLATE 1:** Surface-plot visualization of the density field of the slice $-4000 < Z_{sup} < 4000$ km/sec; the upper panel represents the density field in $z$-space and the lower panel the corresponding 3D-space density field.

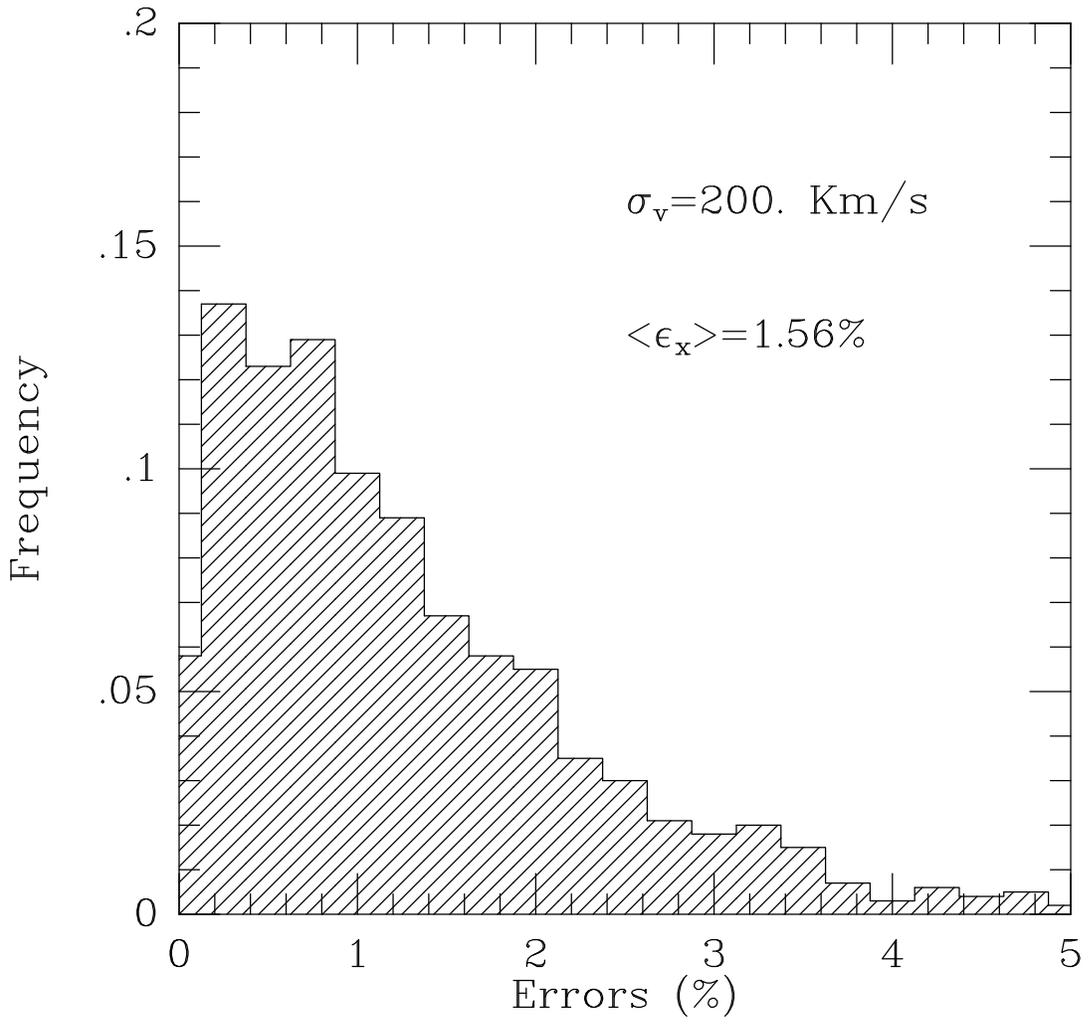

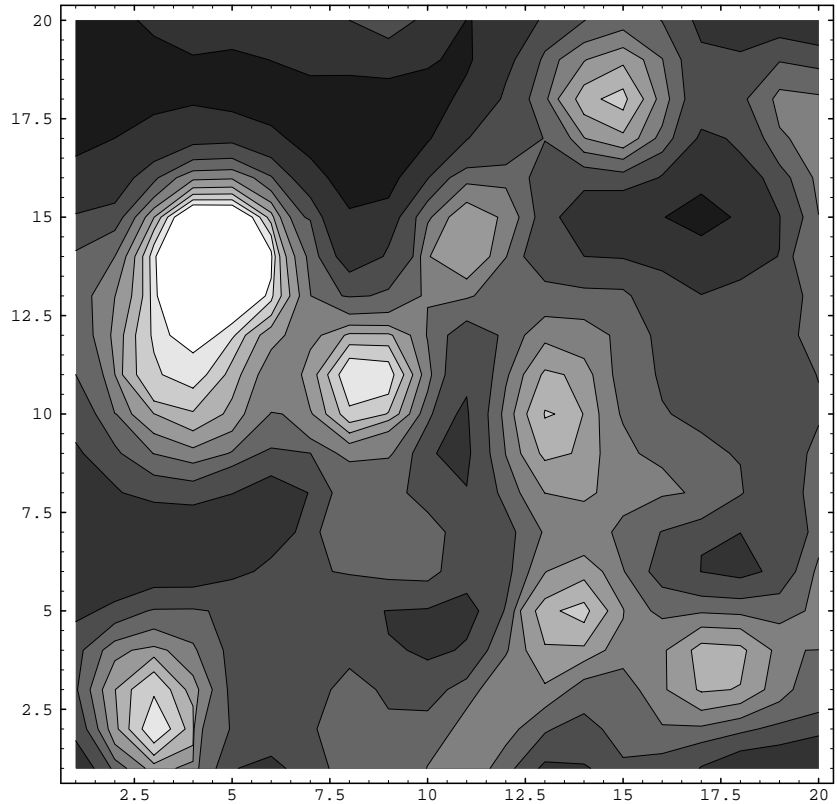

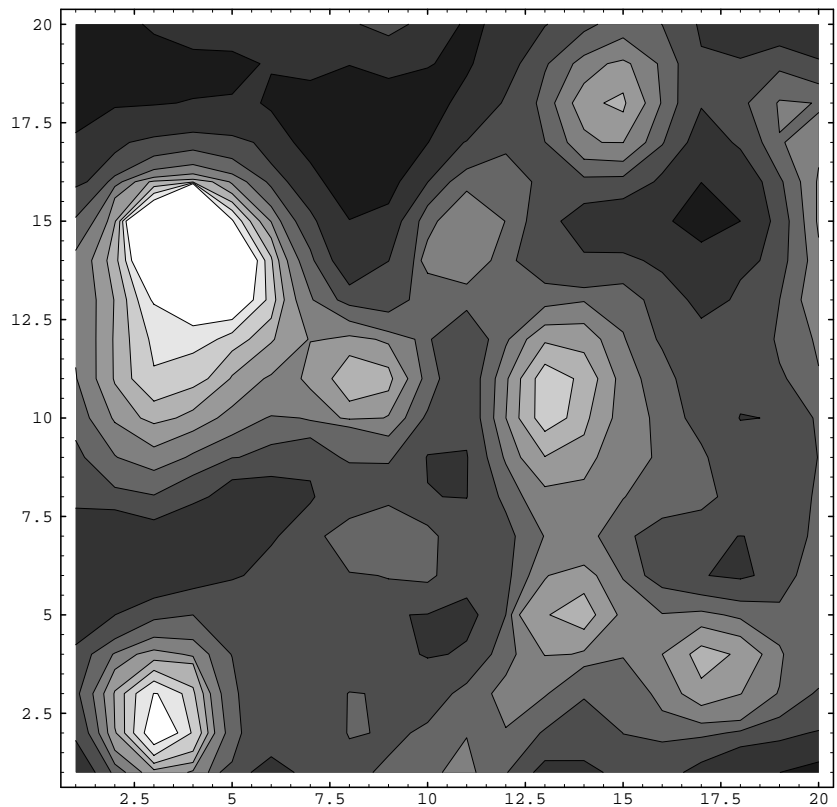

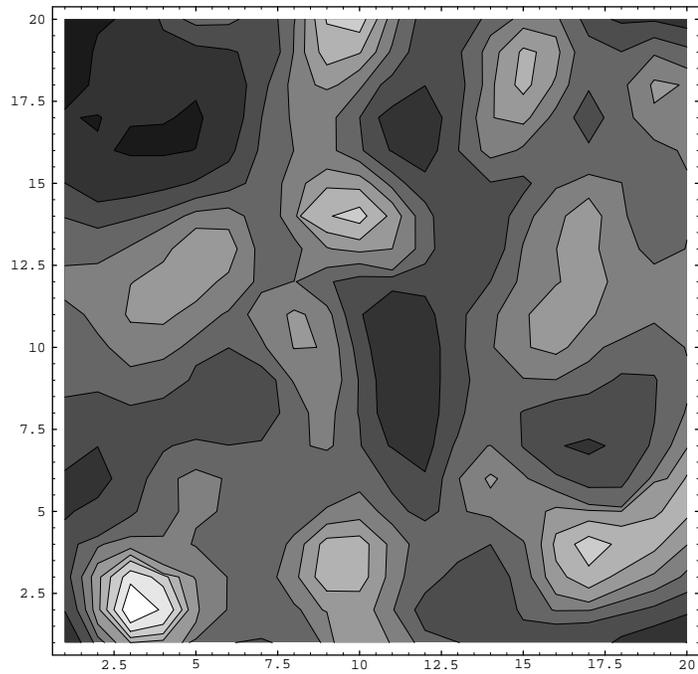
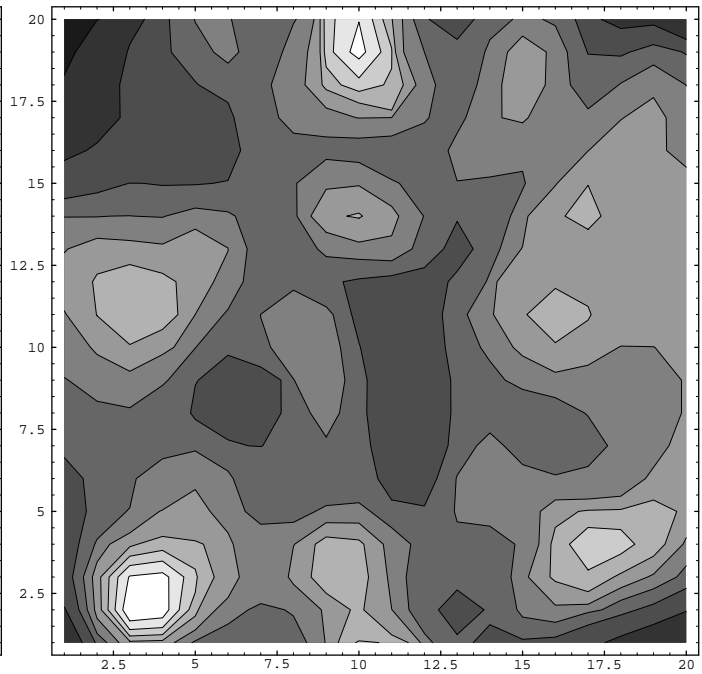
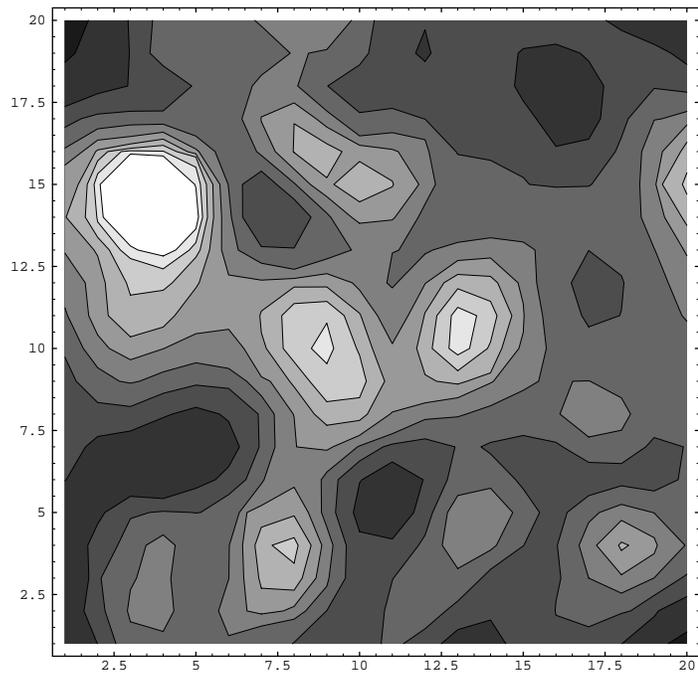
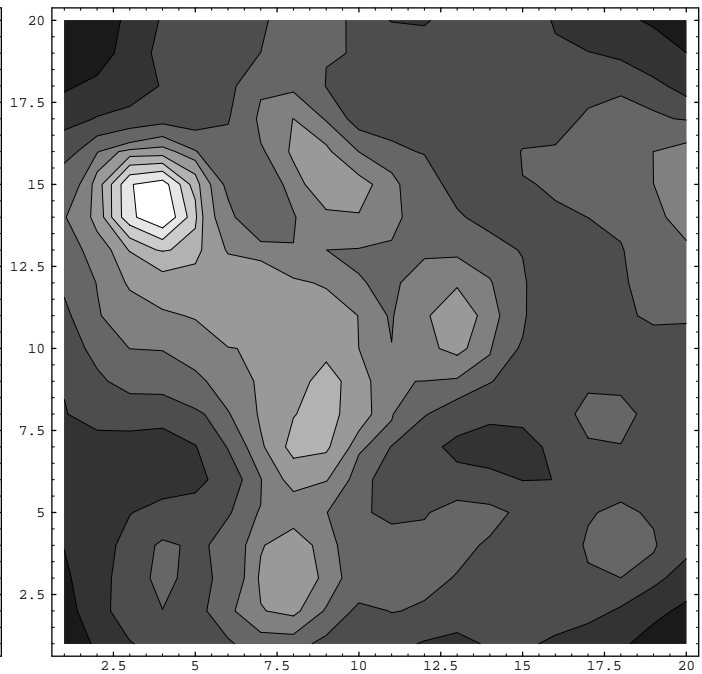

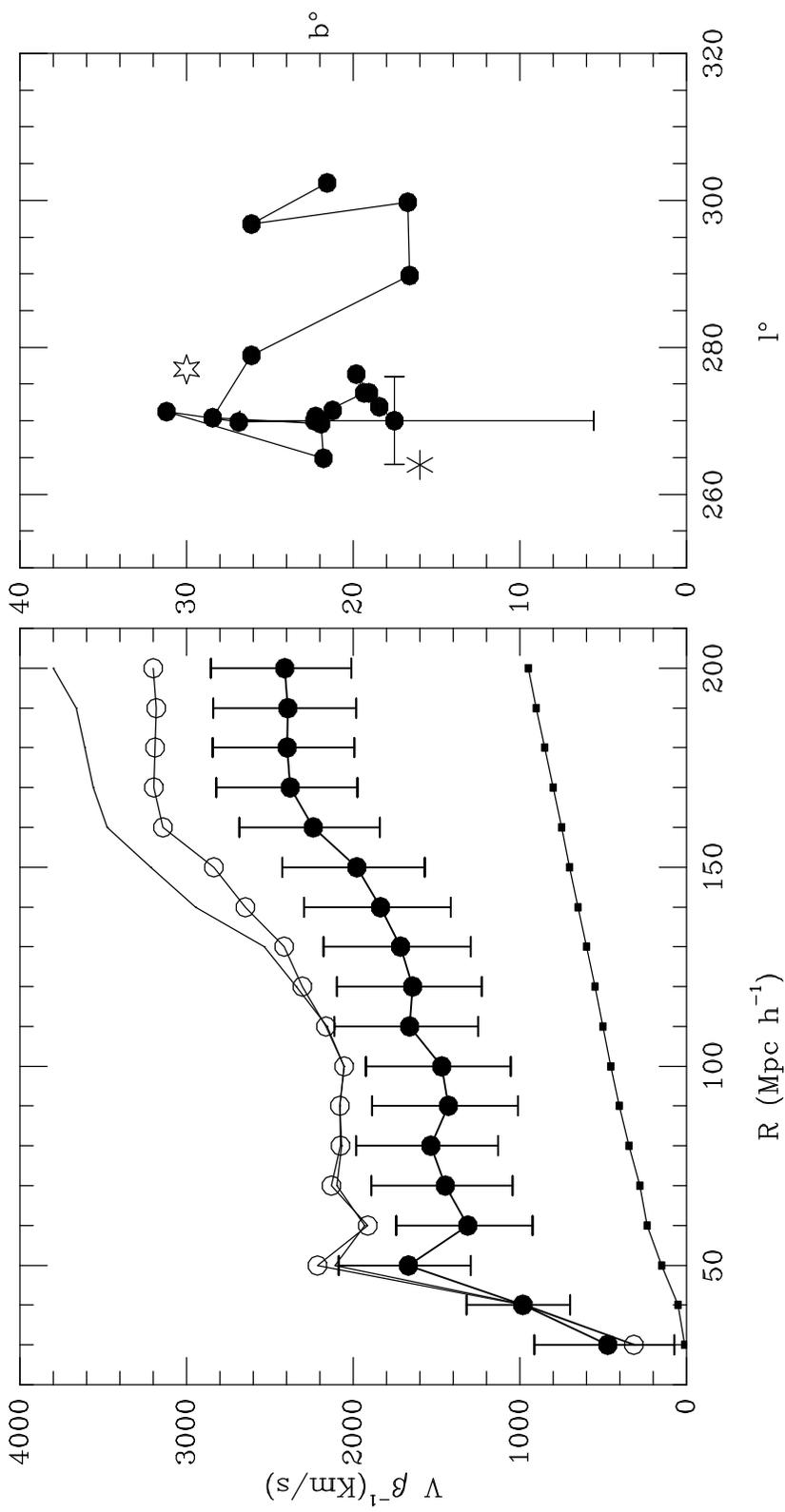

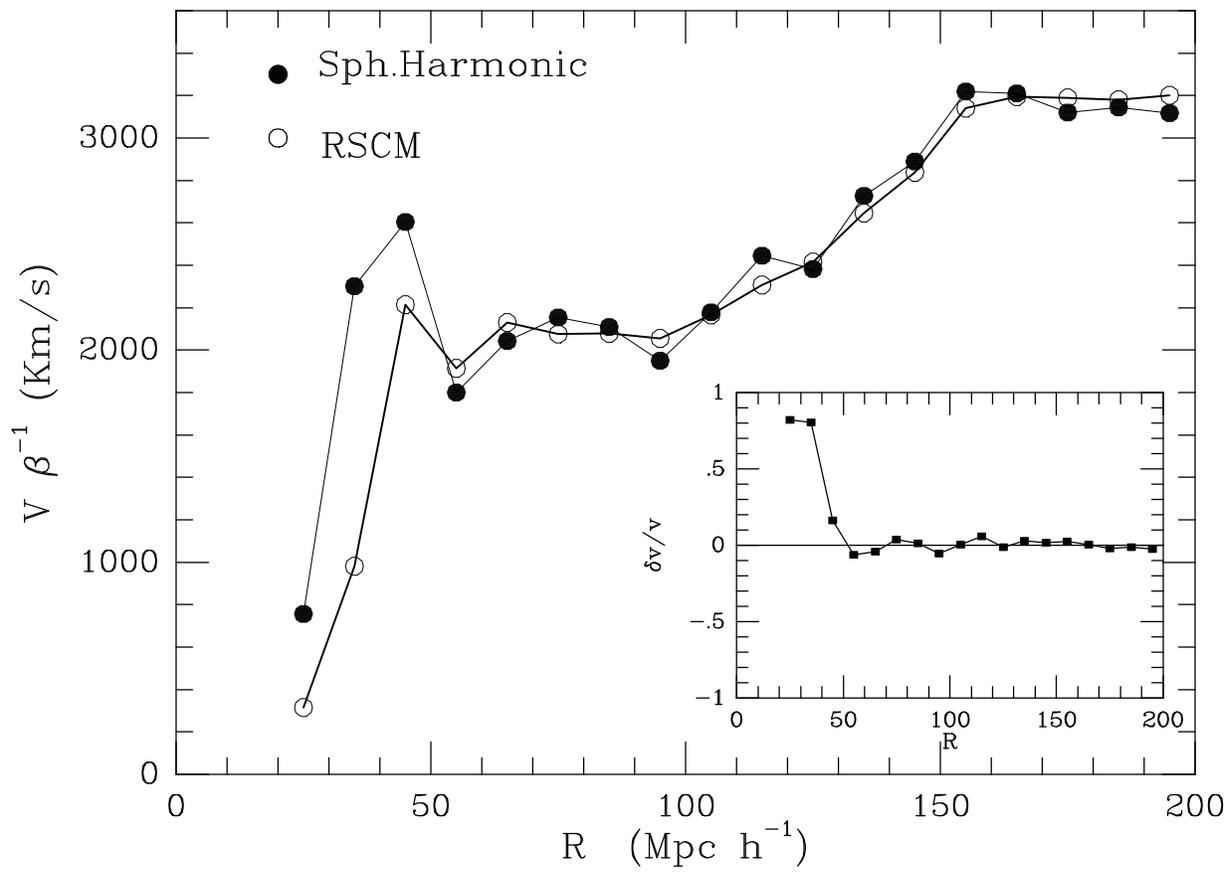

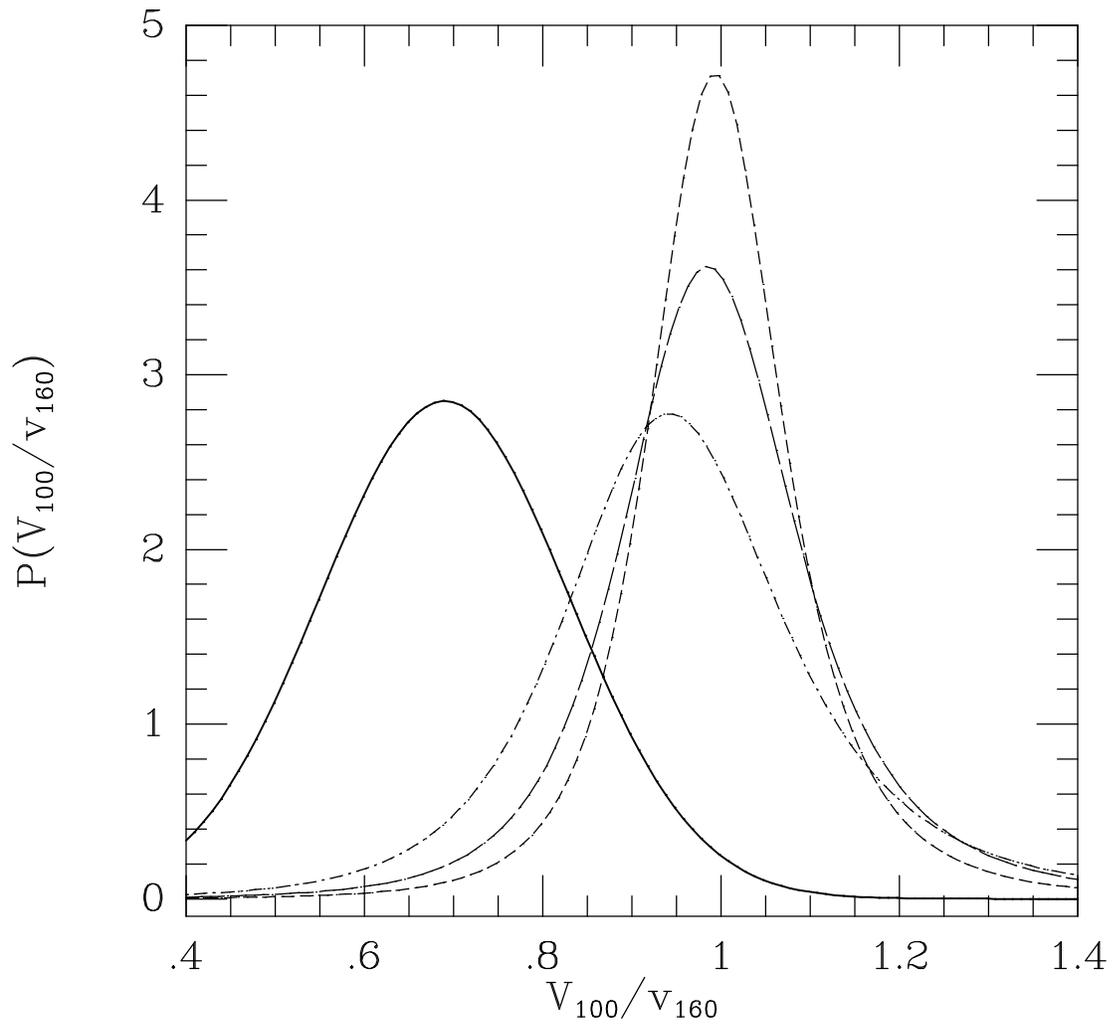

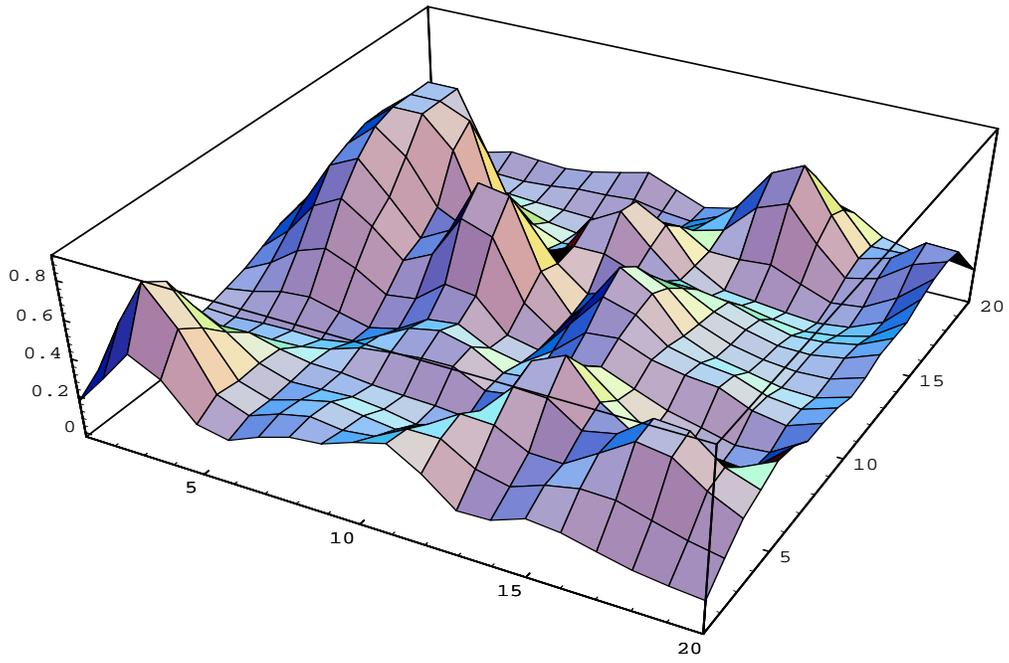

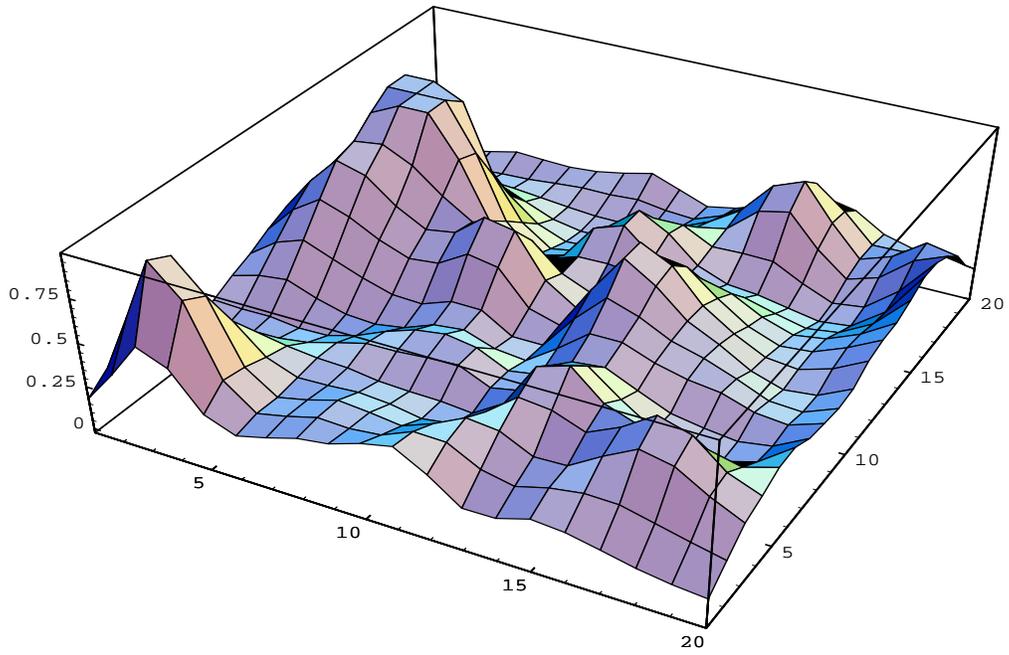

| | | |
|---|---|---|
| Galactic Absorption | **(0.2,0.3)** | (0.2,0.2) |
| Deceleration Parameter $q_o$ | 0.2 | **0.5** |
| Smoothing length $Mpch^{-1}$ | 10 **20** 30 | |
| $W_{rel}(r)$ Binning | **10** | 15 |
| Homogenization Scheme | **Direct** | Inverse |

| Parameters | $\Delta V$ | $\Delta l$ | $\Delta b$ |
|---|---|---|---|
| (0.2,0.3) vs. (0.2,0.2) | $-1 \pm 10\%$ | $+3 \pm 5°$ | $-2 \pm 4°$ |
| $\beta = 0.25$ vs. $\beta = 0.114$ | $-8 \pm 12\%$ | $0 \pm 5°$ | $-1 \pm 4°$ |
| $R_s = 20$ Mpc vs. $R_s = 10$ Mpc | $-3 \pm 11\%$ | $0 \pm 4°$ | $+3 \pm 3°$ |
| $R_s = 20$ Mpc vs. $R_s = 30$ Mpc | $-2 \pm 11\%$ | $+1 \pm 4°$ | $+1 \pm 4°$ |
| bin = 10 vs. bin=15 | $-4 \pm 11\%$ | $+1 \pm 4°$ | $0 \pm 4°$ |
| Direct vs. Inverse | $+4 \pm 11\%$ | $+1 \pm 4°$ | $0 \pm 4°$ |